\documentclass{ws-ijmpcs}

\usepackage[english,german]{babel}

\usepackage[percent]{overpic}
\usepackage{lscape}
\usepackage{amsmath}
\usepackage{bbm}
\usepackage{amssymb}
\usepackage{appendix}
\usepackage{rotating}
\usepackage[percent]{overpic}
\usepackage[usenames, dvipsnames]{color}
\usepackage{longtable}
\usepackage[verbose]{placeins}

\usepackage{theorem}

\begin{document}

\selectlanguage{german}

\markboth{Yannick Wunderlich}
{Studies on a complete experiment for single pseudoscalar meson photoproduction}

%
\catchline{}{}{}{}{}
%

\title{STUDIES ON A COMPLETE EXPERIMENT FOR SINGLE PSEUDOSCALAR MESON PHOTOPRODUCTION
}

\author{Yannick Wunderlich for the CB-ELSA/TAPS collaboration
}

\address{Helmholtz-Institut f"ur Strahlen- und Kernphysik, Universit"at Bonn\\ Nussallee 14-16,
53115 Bonn, 
Germany
\\
wunderlich@hiskp.uni-bonn.de}



\maketitle


\selectlanguage{english}

\begin{abstract}
The complete experiment problem for photoproduction of single pseudoscalar mesons is reviewed briefly. If this problem is investigated in the context of a truncated partial wave analysis, the chance emerges to obtain a unique multipole solution measuring less polarization observables than demanded by the classical solution of the problem that is formulated in terms of spin amplitudes.
\end{abstract}

\ccode{PACS numbers: 11.25.Hf, 123.1K}

\section{Introduction}

The most general expression for the amplitude describing the reaction of photoproduction of single pseudoscalar mesons reads\cite{CGLN}
\begin{align}
  F_{\mathrm{CGLN}} &= i \left( \vec{\sigma} \cdot \hat{\epsilon} \right) F_{1} + \left( \vec{\sigma} \cdot \hat{q} \right) 
  \left[ \vec{\sigma} \cdot \left( \hat{k} \times \hat{\epsilon} \right) \right] F_{2} + i \left( \vec{\sigma} \cdot \hat{k} \right) \left( \hat{q} \cdot \hat{\epsilon} \right) F_{3} \nonumber \\
&\hspace*{9pt}+ i \left( 
   \vec{\sigma} \cdot \hat{q} \right) \left( \hat{q} \cdot \hat{\epsilon} \right) F_{4} \mathrm{.} \label{eq:DefCGLN}
 \end{align}
The four CGLN amplitudes $F_{i} \left(W, \theta  \right)$, which depend on the total energy $W$ and CMS scattering angle $\theta$, completely describe the process. These four amplitudes are accompanied by 16 measurable polarization observables\cite{Sandorfi} at each point in $\left(W, \theta  \right)$. The observables are divided into the four classes: single spin (Group S), beam-target (BT), beam-recoil (BR) and target-recoil (TR). \newline
These facts have lead to the formulation of the so called complete experiment problem\cite{BarDo} (for a more recent publication, see Chiang and Tabakin\cite{ChTab}), which adresses the question of how many and which polarization observables have to be measured in order to uniquely determine the amplitudes. To start with, this problem is purely academic and disregards measurement uncertainties\cite{LotharNStar}.

\section{The Complete Experiment}

Additionally to CGLN amplitudes, helicity amplitudes $H_{i} \left(W, \theta  \right)$ can be defined as well\cite{ChTab}, which are fully equivalent to the former. When written in terms of the $H_{i}$,

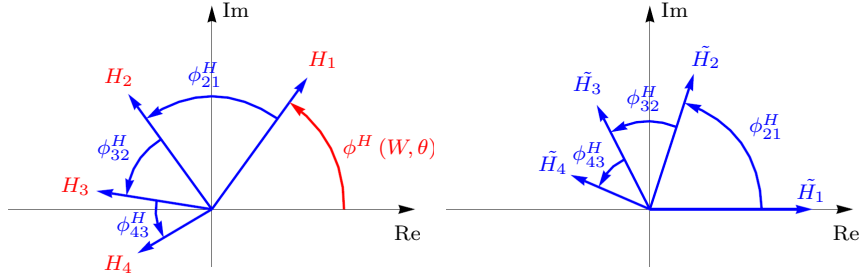
\begin{figure}[ht]
\centering
\begin{overpic}[width=0.45\textwidth]%
      {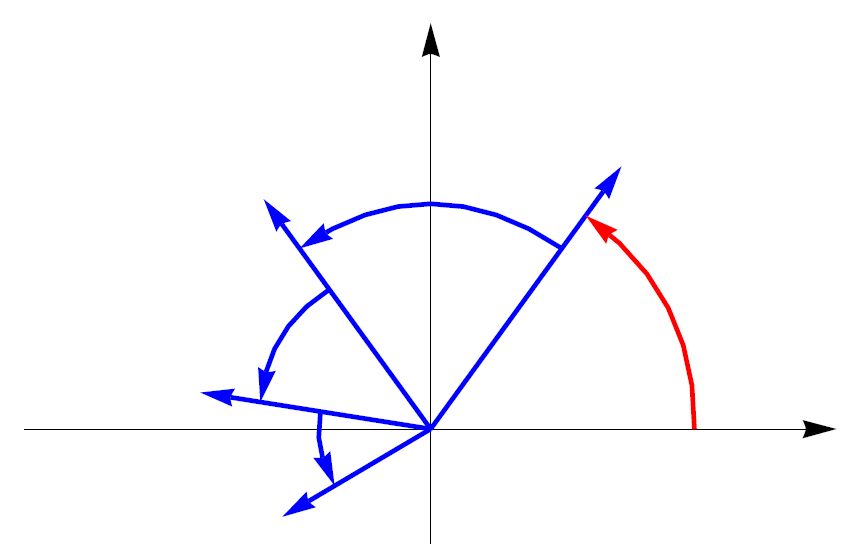}
  \put(92.6,7.0){\begin{footnotesize}\textcolor{black}{Re}\end{footnotesize}}
  \put(52.65,58.75){\begin{footnotesize}\textcolor{black}{Im}\end{footnotesize}}
  \put(80.6,27.0){\begin{footnotesize}\textcolor{red}{$ \phi^{H} \left( W, \theta \right) $}\end{footnotesize}}
  \put(72.6,47.5){\begin{footnotesize}\textcolor{red}{$ H_{1} $}\end{footnotesize}}
  \put(25.1,43.5){\begin{footnotesize}\textcolor{red}{$ H_{2} $}\end{footnotesize}}
  \put(15.0,18.0){\begin{footnotesize}\textcolor{red}{$ H_{3} $}\end{footnotesize}}
  \put(25.0,0.0){\begin{footnotesize}\textcolor{red}{$ H_{4} $}\end{footnotesize}}
  \put(44.6,44.0){\begin{footnotesize}\textcolor{blue}{$\phi^{H}_{21}$}\end{footnotesize}}
  \put(23.5,27.0){\begin{footnotesize}\textcolor{blue}{$\phi^{H}_{32}$}\end{footnotesize}}
  \put(28.0,9.0){\begin{footnotesize}\textcolor{blue}{$\phi^{H}_{43}$}\end{footnotesize}}
\end{overpic}
\begin{overpic}[width=0.45\textwidth]%
      {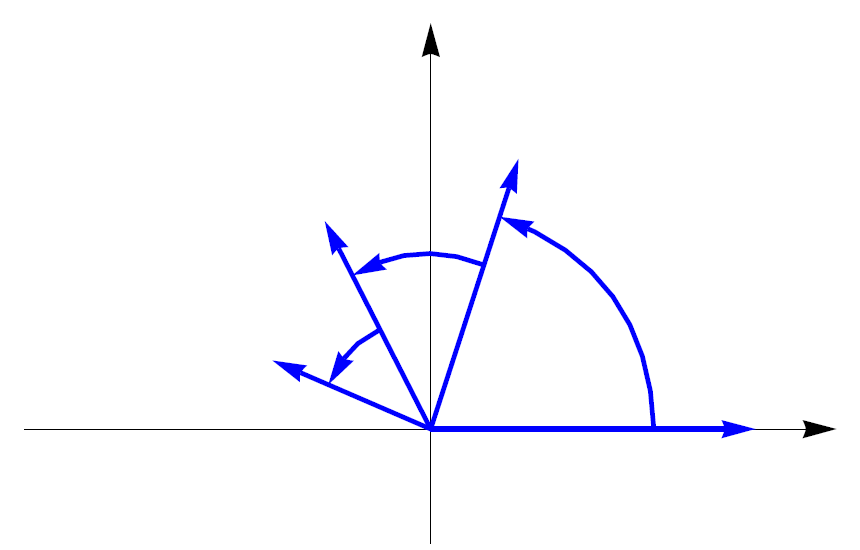}
  \put(92.6,7.0){\begin{footnotesize}\textcolor{black}{Re}\end{footnotesize}}
  \put(52.65,58.75){\begin{footnotesize}\textcolor{black}{Im}\end{footnotesize}}
  \put(84.1,16.5){\begin{footnotesize}\textcolor{blue}{$ \tilde{H_{1}} $}\end{footnotesize}}
  \put(59.6,47.5){\begin{footnotesize}\textcolor{blue}{$ \tilde{H_{2}} $}\end{footnotesize}}
  \put(32.1,41.5){\begin{footnotesize}\textcolor{blue}{$ \tilde{H_{3}} $}\end{footnotesize}}
  \put(24.0,23.0){\begin{footnotesize}\textcolor{blue}{$ \tilde{H_{4}} $}\end{footnotesize}}
  \put(73.1,32.0){\begin{footnotesize}\textcolor{blue}{$\phi^{H}_{21}$}\end{footnotesize}}
  \put(44.6,38.5){\begin{footnotesize}\textcolor{blue}{$\phi^{H}_{32}$}\end{footnotesize}}
  \put(32.0,25.5){\begin{footnotesize}\textcolor{blue}{$\phi^{H}_{43}$}\end{footnotesize}}
\end{overpic}
\caption{The left hand side shows unconstrained helicity amplitudes $H_{i}$. The right hand side shows constrained amplitudes $\tilde{H}_{i}$ that can be determined solely from a complete experiment.}
\label{fig:CompareCompExAmplitudesToActualAmplitudes}
\end{figure}
the profile functions $\check{\Omega}^{\alpha}$ defining the 16 polarization observables $\Omega^{\alpha}$ via $\check{\Omega}^{\alpha} = \Omega^{\alpha} \check{\Omega}^{1}$ become bilinear hermitean forms\cite{ChTab}
\begin{equation}
\check{\Omega}^{\alpha} = \frac{1}{2} \sum \limits_{i,j=1}^{4} H_{i}^{\ast} \Gamma^{\alpha}_{ij} H_{j} = \frac{1}{2} \left< H \right| \Gamma^{\alpha} \left| H \right> \mathrm{,} \quad \alpha = 1,\ldots,16 \mathrm{,} \label{eq:BHPForm}
\end{equation}
with $\left\{\Gamma^{\alpha}\right\}$ being a set of 16 hermitean unitary matrices. It is possible to formally invert these bilinear equations to yield\cite{ChTab}
\begin{equation}
 H_{i}^{\ast} H_{j} = \frac{1}{2} \sum \limits_{\alpha}^{} \left(\Gamma^{\alpha}_{ij}\right)^{\ast} \check{\Omega}^{\alpha}  \mathrm{.} \label{eq:H-Products}
\end{equation}
The bilinear products $H_{i}^{\ast} H_{j}$ allow for the extraction of moduli $\left| H_{i} \right|$ as well as relative phases $\phi^{H}_{ij}$ of the helicity amplitudes (the same holds for photoproduction amplitudes defined in all other bases). \newline
It is a central statement of Chiang and Tabakin\cite{ChTab} that the amplitudes can be determined from complete sets of at least 8 observables. These sets have to contain all four Group S observables $\left(\frac{d \sigma}{d \Omega}\right)_{0}$, $\Sigma$, $T$ and $P$. The remaining four measurements must not belong to the same class of double polarization measurements. Additionally, no more than 2 observables are to be picked from the same double polarization class. The complete sets are listed\cite{ChTab}. \newline
Amplitudes can only be determined from a complete experiment up to an energy and angular dependent overall phase $ \phi^{H} \left( W, \theta \right) $, see Figure \ref{fig:CompareCompExAmplitudesToActualAmplitudes}. Partial waves (multipoles) on the other hand can only be extracted from full production amplitudes by means of projection equations that involve angular integrations\cite{JSBall}. These two facts yield the result that helicity amplitudes extracted from a complete experiment up to an overall phase do not give access to partial waves (the same holds for amplitudes in all other bases). A solution to this problem consists of the performance of a truncated partial wave analysis (TPWA), which is discussed in the next section.

\section{Truncated Partial Wave Analysis (TPWA)}

\begin{figure}[hb]
\begin{overpic}[width=0.8625\textwidth]%
      {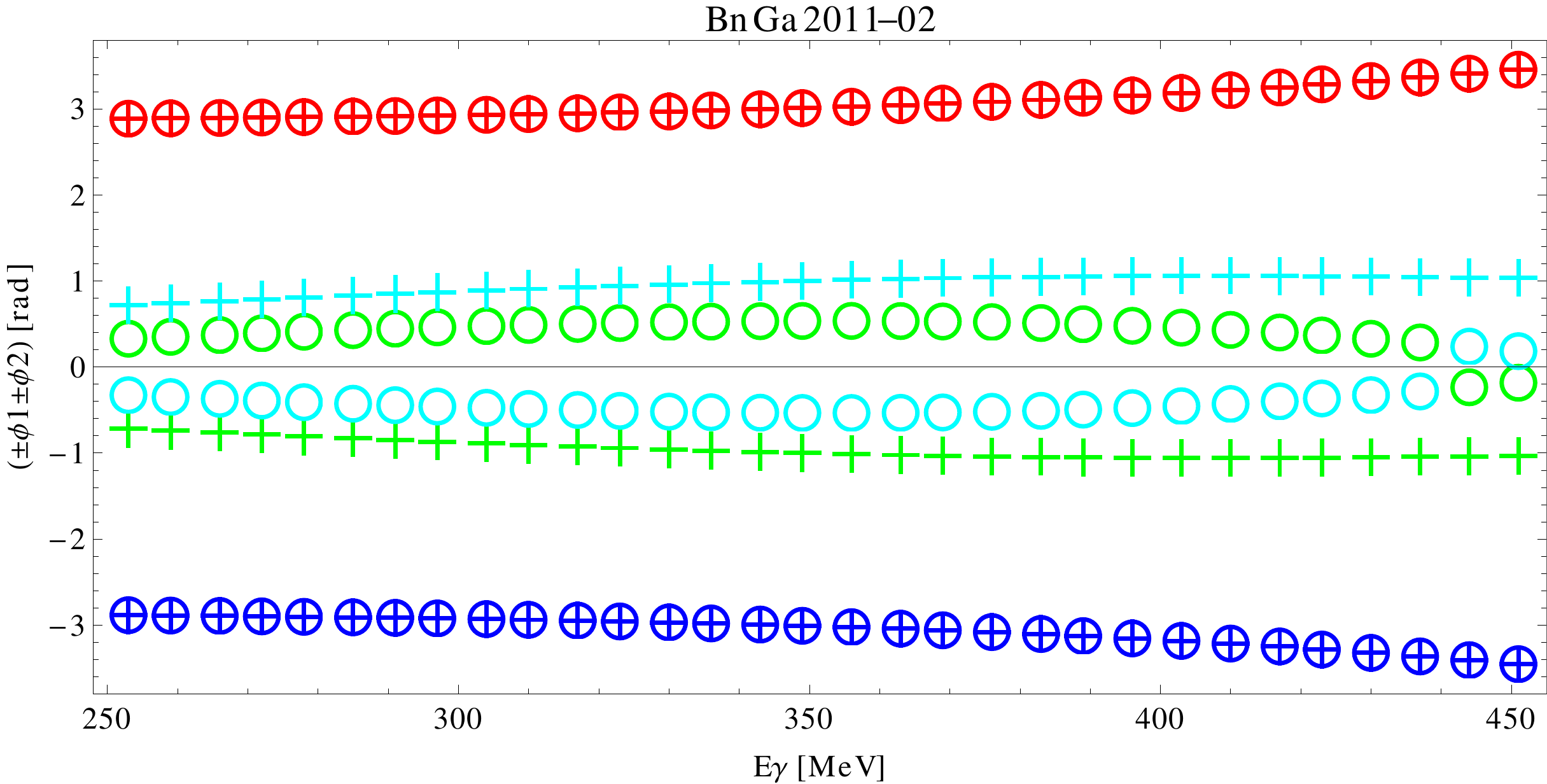}
         \put(101.6,7){\begin{footnotesize}\textcolor{blue}{$\updownarrow$ solution 1}\end{footnotesize}}
         \put(101.6,45.5){\begin{footnotesize}\textcolor{red}{$\updownarrow$ solution 2}\end{footnotesize}}
         \put(101.6,30){\begin{footnotesize}\textcolor{green}{$\updownarrow$ solution 3}\end{footnotesize}}
         \put(101.6,23){\begin{footnotesize}\textcolor{cyan}{$\updownarrow$ solution 4}\end{footnotesize}}
 \end{overpic}
\caption{The ambiguity diagram for $\ell_{\mathrm{max}}=1$, generated using the solution BnGa 2011-02. The solutions mentioned in the main text are indicated. A similar diagram is given by Omelaenko$^{7}$.}
\label{fig:BnGaAmbiguityDiagram}
\end{figure}

\begin{figure}[ht]
\centering
\begin{overpic}[width=0.393\textwidth]%
      {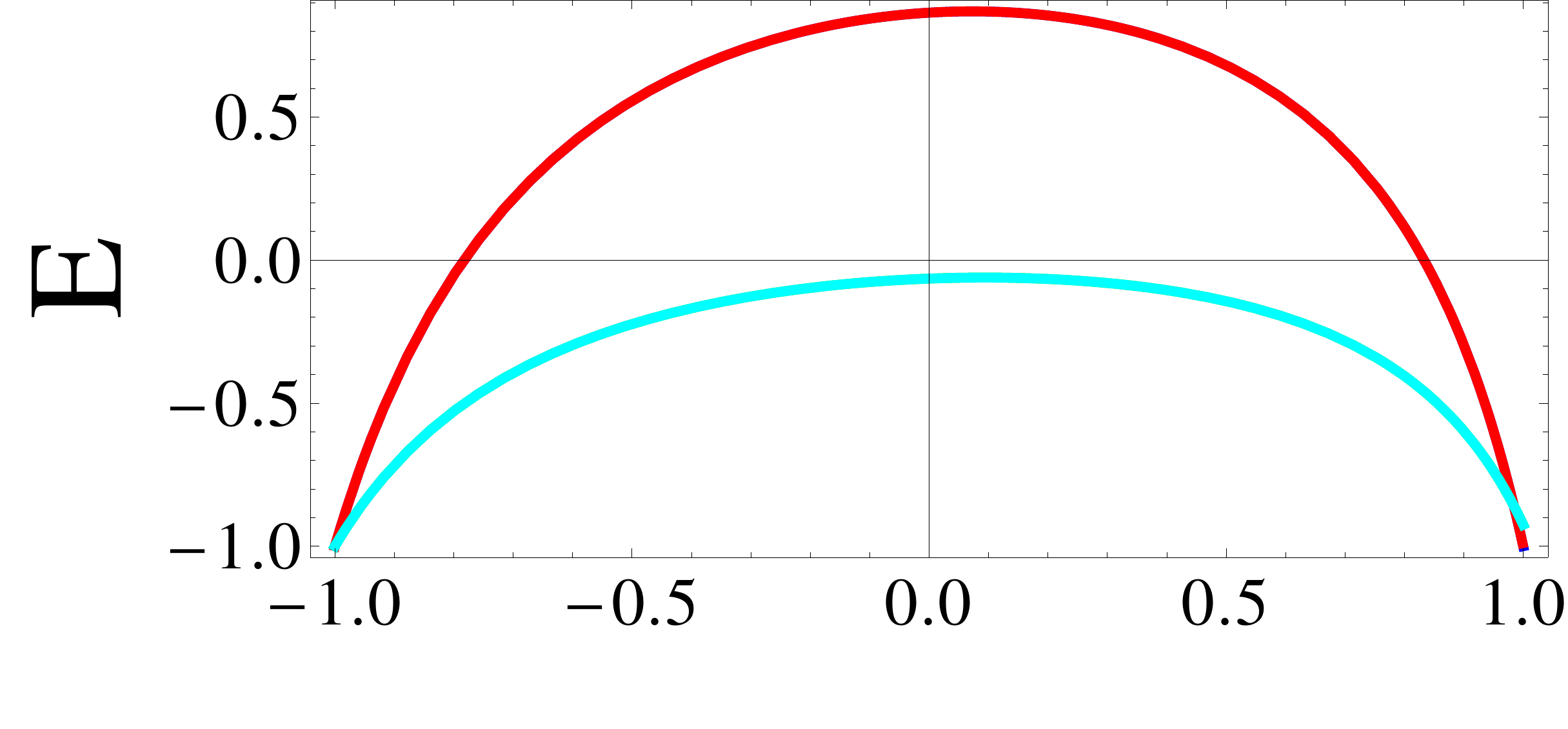}
\end{overpic}
\begin{overpic}[width=0.393\textwidth]%
      {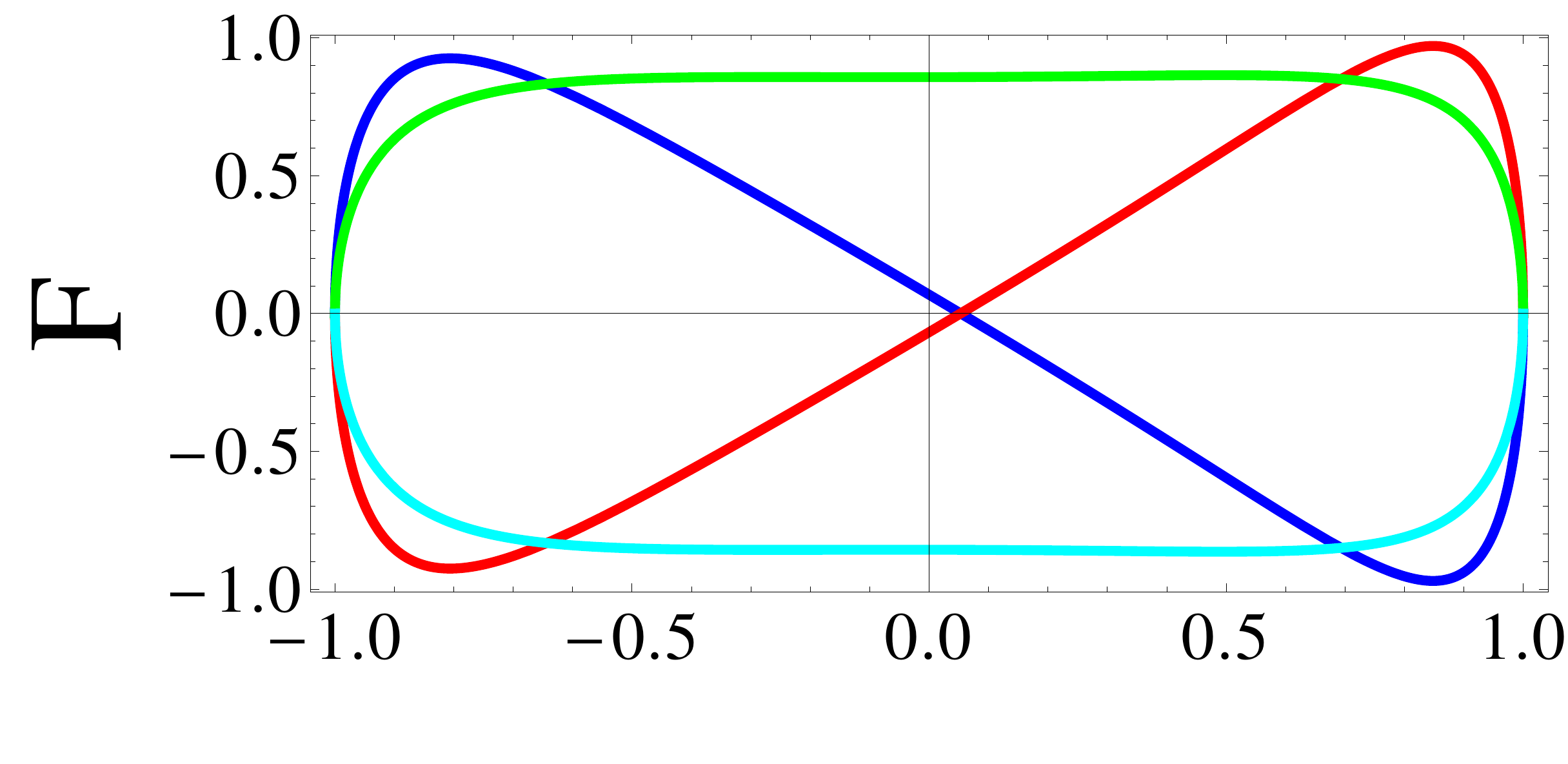}
\end{overpic} \\
\begin{overpic}[width=0.393\textwidth]%
      {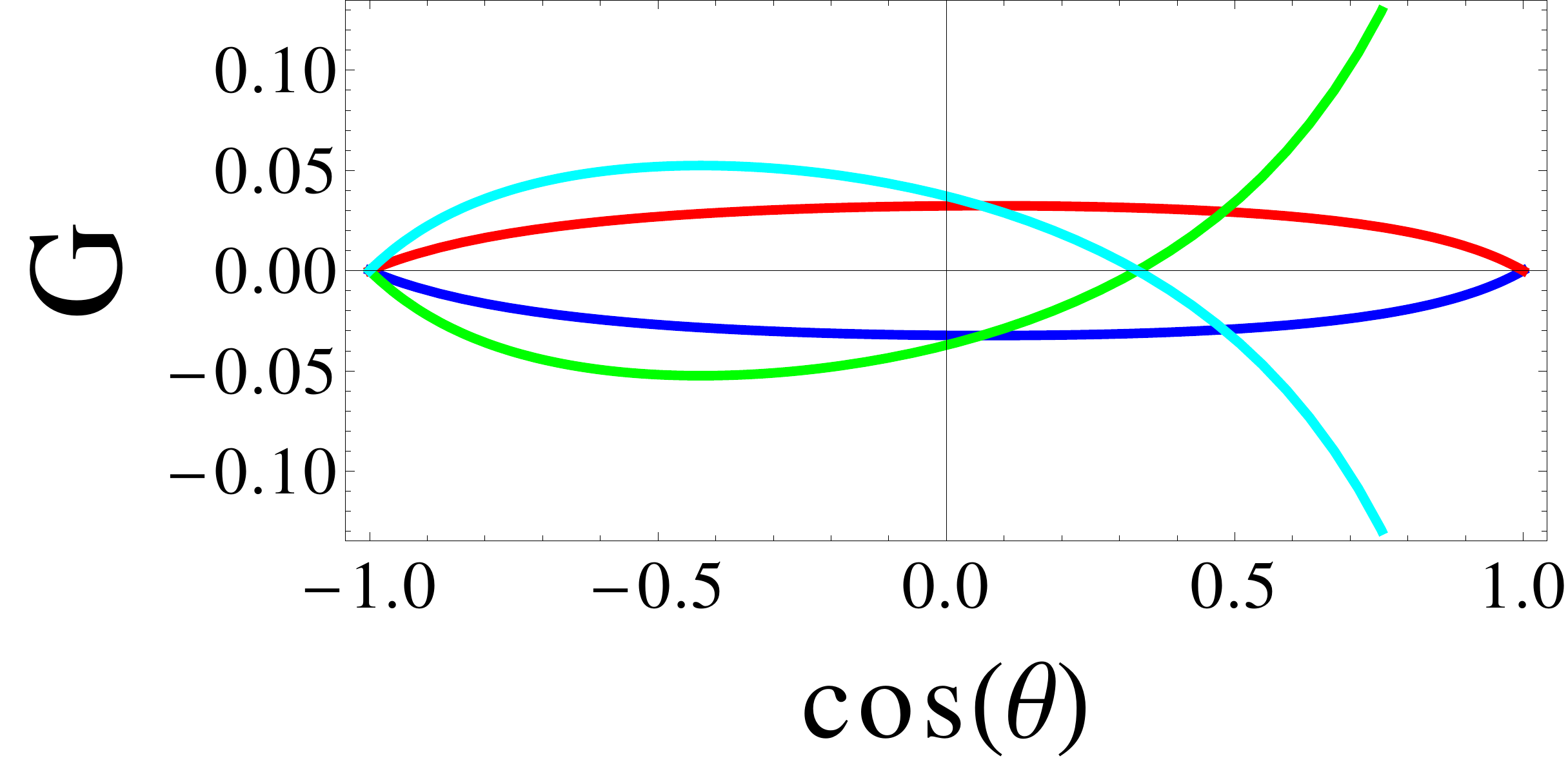}
\end{overpic}
\begin{overpic}[width=0.393\textwidth]%
      {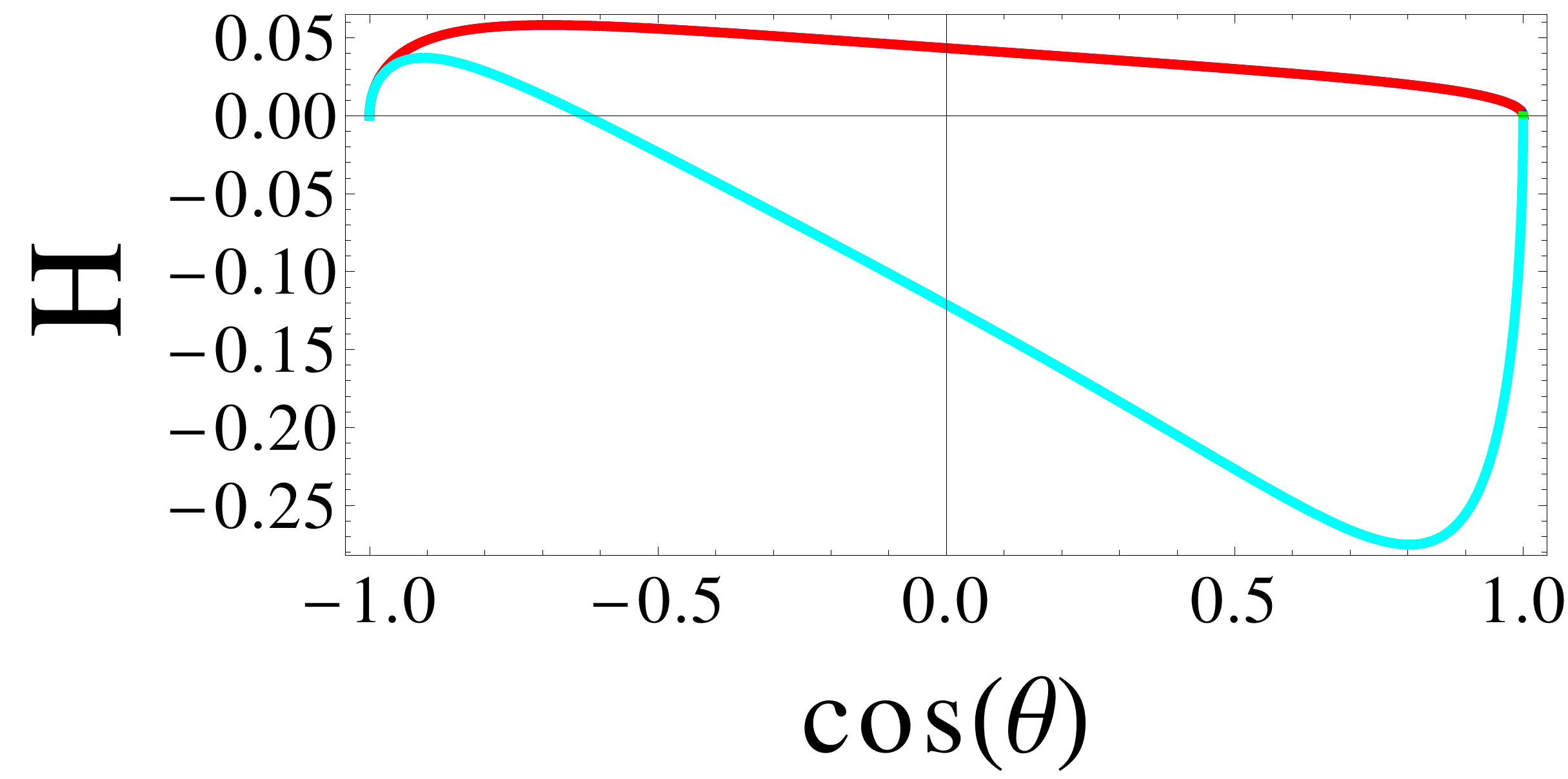}
\end{overpic}
\caption{Preliminary plot of $\left\{E,\hspace*{1pt} F,\hspace*{1pt} G,\hspace*{1pt} H\right\}$ ($E_{\gamma} = 253 \hspace*{2pt} \mathrm{MeV}$), for all four solutions.}
\label{fig:BnGaCompareObservables}
\end{figure}

The multipole expansion of the CGLN amplitudes is well known\cite{Sandorfi}. Once this expansion is truncated at some finite angular momentum $\ell_{\mathrm{max}}$ and this truncation is inserted into the polarization observables, the latter acquire an angular parametrization that is summarized by the following equations\cite{LotharNStar}
\begin{align}
\check{\Omega}^{\alpha} \left( W, \theta \right) &= \sin^{\beta_{\alpha}} \theta \sum \limits_{k = 0}^{2 \ell_{\mathrm{max}} + \gamma_{\alpha}} a_{k}^{\alpha} \left( W \right) \cos^{k} \theta \mathrm{,}  \label{eq:LowEParametrization} \\
a_{k}^{\alpha} \left( W \right) &= \sum \limits_{\ell, \ell' = 0}^{\ell_{\mathrm{max}}} \sum \limits_{\kappa, \kappa' = 1}^{4} \mathcal{C}_{\ell, \ell'}^{\kappa, \kappa'} \mathcal{M}^{\ast}_{\ell,\kappa} \left( W \right) \mathcal{M}_{\ell', \kappa'} \left( W \right) \mathrm{.} \label{eq:BilinearCoeffs}
\end{align}
%
Here, the profile functions $\check{\Omega}^{\alpha}$ are expanded in finite powers of $\cos \theta$, with real expansion cofficients $a_{k}^{\alpha}$ and energy dependent complex multipoles $\mathcal{M}_{\ell,\kappa} \left( W \right)$. $\beta_{\alpha}$ and $\gamma_{\alpha}$ are constants depending on the observable under consideration\cite{LotharNStar}. For every finite order in $\ell_{\mathrm{max}} \geq 1$, $4 \ell_{\mathrm{max}}$ complex multipoles appear in the expansion. 
The question of which ambiguities exist for the Group S observables in a TPWA was discussed by Omelaenko\cite{Omelaenko} using ideas by Gersten\cite{Gersten}. There, the $4 \ell_{\mathrm{max}}$ complex multipoles are exchanged for an equivalent set of complex variables $\alpha_{j}$ and $\beta_{k}$ that are ideally suited to determine all ambiguities of the Group S observables:
\begin{equation}
\left\{ \mathcal{M}_{\ell,\kappa} \left( W \right) \right\} \leftrightarrow \left\{ \alpha_{j}, \hspace*{1pt} \beta_{k} \right\} = \left\{ \left|\alpha_{j}\right| e^{i \phi_{j}} , \hspace*{1pt} \left|\beta_{k}\right| e^{i \psi_{k}} \right\} \mathrm{,} \quad j,k = 1,\ldots,2 \ell_{\mathrm{max}} \mathrm{.}
\end{equation}
Once an already existing multipole solution is used as a numerical input, the question of the arising ambiguities can be discussed in the form of a diagram\cite{Omelaenko}. Using a Bonn Gatchina\cite{BoGa} solution as input, an ambiguity diagram is drawn in Figure \ref{fig:BnGaAmbiguityDiagram} for the simple case $\ell_{\mathrm{max}}=1$. The different cases of linear combinations of phases $\pm \phi_{1} \pm \phi_{2} \equiv \circ$ and $\pm \psi_{1} \pm \psi_{2} \equiv +$ are drawn in different colors. Once circles and crosses of any color almost coincide, an ambiguity of the Group S observables appears. This rule allows for the deduction of four ambiguous solutions (Fig. \ref{fig:BnGaAmbiguityDiagram}). \newline
Once BT observables $\left\{E,\hspace*{1pt} F,\hspace*{1pt} G,\hspace*{1pt} H\right\}$ are plotted for those four solutions (Fig. \ref{fig:BnGaCompareObservables}), some solutions for the observables $E$ and $H$ cannot be distinguished.
\newpage
$G$ and $F$ are different for all four solutions. This investigation therefore allows the postulation of sets of only 5 observables that should facilitate an unambiguous extraction of multipoles in a TPWA. This statement is consistent with Omelaenko\cite{Omelaenko} as well as Grushin\cite{Grushin}. The reformulation of the TPWA first worked out by Omelaenko\cite{Omelaenko} is planned to be revisited in a work that has yet to be published\cite{OmelaenkoRevisited}.

\section{Summary}

A set of four spin amplitudes describing pseudoscalar meson photoproduction is determined theoretically by measuring certain sets of 8 polarization observables. The undeterminable overall phase of those amplitudes denies access to partial waves. A way out of this issue consists of a truncated partial wave expansion (TPWA). First studies suggest that the multipoles of this expansion can be obtained by using even less than 8 observables. \newline
This work was supported by the \textit{Deutsche Forschungsgemeinschaft} within SFB/TR16.

\selectlanguage{english}


\end{document}